\documentclass[sigconf]{acmart}

\usepackage{array,multirow,graphicx}
\usepackage{subfigure}
\usepackage{booktabs} 
\usepackage{amsmath}
\usepackage{mathtools}
\usepackage[linesnumbered,lined,boxed,commentsnumbered]{algorithm2e}

\begin{document}
\title{Prediction of next career moves from scientific profiles}

\author{Charlotte James}
\orcid{1234-5678-9012}
\affiliation{
  \institution{
  Department of Engineering Mathematics\\University of Bristol}
  \city{Bristol} 
  \state{UK} 
}
\email{charlotte.james@bristol.ac.uk}

\author{Luca Pappalardo}
\orcid{0000-0002-1547-6007}
\affiliation{
  \institution{
  Institute of Information Science and Technologies (ISTI)\\
  National Research Council (CNR)}
  \city{Pisa} 
  \state{Italy} 
 }
\email{luca.pappalardo@isti.cnr.it}

\author{Alina S\^irbu}
\affiliation{
  \institution{Department of Computer Science\\University of Pisa}
  \city{Pisa} 
  \country{Italy}}
\email{alina.sirbu@unipi.it}

\author{Filippo Simini}
\affiliation{
  \institution{Department of Engineering Mathematics\\University of Bristol}
  \city{Bristol}
  \country{UK}
}
\email{f.simini@bristol.ac.uk}

\renewcommand{\shortauthors}{C. James, L. Pappalardo, A. S\^irbu, F. Simini}

\begin{abstract}
Changing institution is a scientist's key career decision, which plays an important role in education, scientific productivity, and the
generation of scientific knowledge. Yet, our understanding of the factors influencing a relocation decision is very limited. 
In this paper we investigate how the scientific profile of a scientist determines their decision to move (i.e., change institution). To this aim, we describe a scientist's profile by three main aspects: the scientist's recent scientific career, the quality of their scientific environment and the structure of their scientific collaboration network. We then design and implement a two-stage predictive model: first, we use data mining to predict which researcher will move in the next year on the basis of their scientific profile; second we predict which institution they will choose by using a novel social-gravity model, an adaptation of the traditional gravity model of human mobility. Experiments on a massive dataset of scientific publications show that our approach performs well in both the stages, resulting in a 85\% reduction of the prediction error with respect to the state-of-the-art approaches.
\end{abstract}

\maketitle

\section{Introduction}
Changing institution is an integral part of academic life and a scientist's key career decision, which can potentially play an important role in education, scientific productivity, and the
generation of scientific knowledge. Scientific migration allows scientists to find environments where they are more effective in doing their research and contribute to the success of research institutions. Yet, our understanding of the factors influencing a relocation decision, such as the scientific profile of a scientist or the quality of their scientific environment, is limited. Furthermore, while the role of social relationships is known to influence human activity in several contexts \cite{cho2011friendship,wang2013quantifying}, it is not clear what the contribution is of a scientist's collaboration network on their decision to relocate.
Previous research on the migration patterns of scientists is limited to the analysis of large-scale surveys on country-level movements~\cite{azoulay2017mobility,appelt2015factors}, or to the investigation of the effects of changing institution on scientific production and citations~\cite{deville2014career,sugimoto2017scientists}. 

In this work we investigate the other perspective, i.e., how the scientific profile of a scientist influences their decision to move, based on a large dataset of publications from the American Physical Society (APS), consisting of all the publications in APS journals from 1950 to 2009 -- 60,000 scientists, 3,500 institutions and 360,000 articles. Our main objective is to predict where a scientist will move in the next year. 
We approach this problem by dividing it into two subproblems, constructing thus a \emph{two-stage predictive model}. We first predict, using data mining techniques, which researcher in the APS database will move (i.e., change institution) in the next year. At this step we describe a scientist's profile as a multidimensional vector of variables describing three main aspects: the scientist's recent scientific career, the quality of their scientific environment and the structure of their scientific collaboration network. From the constructed predictive model, we also identify the main factors influencing scientific migration. Secondly, for those researchers who are predicted to move, we predict which institution they will choose using a novel \emph{social-gravity} model, an adaptation of the traditional gravity model of human mobility to include the factors identified at the first step. 

Our experiments on the APS dataset reveal two crucial results: (i) a scientist's \emph{xenophilia}, i.e., their tendency to collaborate with scientists at external institutions, is strongly correlated with their decision to migrate; (ii) our approach performs well in both the prediction phases, with the proposed social-gravity model producing a 85\% reduction of the prediction error with respect to the state-of-the-art gravity model.\footnote{The data and the related code to reproduce the research will be made available in the camera-ready version of the paper.} 

This work provides several contributions to the scientific community. First and foremost, we build a novel model of scientific migration that combines data mining with a customised model of human mobility. 
Furthermore, we provide insight, based on data, into the factors affecting the decision to change institution and those involved in choosing the next employer. Our predictive model can help institutions and governments understand scientific mobility and implement policies to attract the best scientists or prevent their departure, hence improving the quality of research. At the same time this type of predictions can facilitate the construction of services that recommend new jobs to scientists based on their profile, or help search committees seek successful candidates for their research jobs.

The rest of the paper is organised as follows. In the next section we discuss related work on scientific migration. In Section~\ref{sec:def} we formalise our problem by defining career trajectories and scientific profiles. Section~\ref{sec:methods} describes our data, the analysis performed, and the results of each of our prediction phases. We conclude the paper with a discussion in Section~\ref{sec:conclusions}. 

\section{Related Work}

Various aspects of scientific profiles have been analysed in recent years, with publication data being central in the process. One line of research is on collaboration networks, which brought important insights and results in the field of complex network theory, and its applications on analysing real-world networks~\cite{perra2012activity,colizza2006detecting,newman2001structure}. A different aspect is the evaluation of the productivity of scientists, relevant for career progression in academia, with multiple performance indices being proposed over the years~\cite{sinatra2016quantifying,wang2013quantifying}. 
Despite their importance to scientific productivity and education, little effort has been put in the literature on understanding how scientists make career moves. The studies proposed in the literature in this regard can be grouped into three strands of research.

A first strand focuses on large-scale surveys on country-level movements and reveal long-term cultural and economic priorities \cite{auriol2010careers, vannoorden2012global, moed2013studying}. Appelt et al. \cite{appelt2015factors} use a gravity-based empirical framework to investigate the factors that influence the international mobility of scientists, finding that geographic distance, as well as socio-economic disparities and scientific proximity, negatively correlate with the mobility of scientists between two countries. Azoulay et al. \cite{azoulay2017mobility} investigate the professional and personal determinants of the decision to relocate to a new institution, finding that scientists are more likely to move when they are highly productive and when 
their local collaborators are fewer and less accomplished than their distant collaborators, 
while they find it costly to disrupt the social networks of their children. 

A second strand of research focuses on understanding the impact of a scientist's relocation to their scientific impact. Analyzing the relocations and the scientific performance of scientists, Deville et al. \cite{deville2014career} find that while moves from elite to lower-rank institutions lead to a moderate decrease in scientific performance, moves to elite institutions does not necessarily result in subsequent performance gain. Sinatra et al. \cite{sinatra2017quantifying} offer empirical evidence that scientific impact is randomly distributed within the sequence of papers published by an individual during the scientific career, implying that temporal changes in impact can be explained by temporal changes in productivity or luck.  
Sugimoto \cite{sugimoto2017scientists} analyzes the migration traces of scientists extracted from Web of Science and reveals that, regardless the nation of origin, scientists who relocate are more highly cited than their non-moving counterparts. 

In the context of studying labor mobility, the availability of massive datasets of individuals' career path fostered works on predicting individuals' next jobs (outside the academia). Paparrizos et al. \cite{paparrizos2011machine} build a system to recommend new jobs to people who are seeking a job, using all their past job transitions as well as their employees data. They train a machine learning model to show that job transitions can be accurately predicted, significantly improving over a baseline that always predicts the most frequent institution in the data. Recently, Li et al. \cite{li2017nemo} propose a system to predict next career moves based on profile context matching and career path mining from a real-world LinkedIn dataset. They show that their system can predict future career moves, revealing interesting insights in micro-level labor mobility.

Our work is placed on the line of conjunction of the aforementioned strands of research. We explore the characteristics of scientific performance which most affect the decision to relocate, focusing on aspects that have not been investigated yet in the literature, such as a scientist's propensity to collaborate with external institutions and their relations with the peer environment. Moreover, we propose an algorithm to predict next career move which is tailored for scientists, hence considering science-specific features and the distance between scientific institutions. From a methodological point of view, our work provides a novel solution to the next (scientific) career move problem, as we combine data mining predictive models with global generative migration models.

\section{Definitions}
In this section we introduce the concept of career trajectory (Section \ref{sec:career_traj}), how we define a scientific profile (Section \ref{sec:profile}) and formalize the problem of next institution prediction (Section \ref{sec:problem}).

\label{sec:def}
\subsection{Career trajectory}
\label{sec:career_traj}
A career trajectory indicates the time-ordered sequence of institutions the scientist worked at during a given time window. 
Formally, we define a scientist $u$'s career trajectory as a sequence of $n$ tuples: $$T(u) = \langle (t_1, i_1), \dots, (t_n, i_n) \rangle,$$ where $t_j = 1, \dots, n$ is a timestamp (year), $\forall_{1 \le j < n} t_j < t_{j + 1}$ and $i_j$ is the scientist's institution (i.e., their affiliation) at time $t_j$, with $i_j \neq i_{j - 1}$. 
Two consecutive affiliations in a career trajectory indicate a \emph{move}, i.e., that the scientists moved from an institution to another. 
A move in $T(u)$ is formally defined as a pair of consecutive tuples in $T(u)$. For example, 
%
% $T(u) = \langle (42.054, -87.676, 1968), (53.551, 9.99, 1973) \rangle$ 
$T(u) = \langle (1968, \text{Evanston}), (1973,\text{Hamburg}) \rangle$ 
is a career trajectory indicating that scientist $u$ moved from Evanston, Illinois %(whose coordinates are 42.054 and -87.676) 
to Hamburg, Germany %(with coordinates 53.551 and 9.99) 
in year 1973. 

\begin{figure}
\centering     
\includegraphics[scale=0.40]{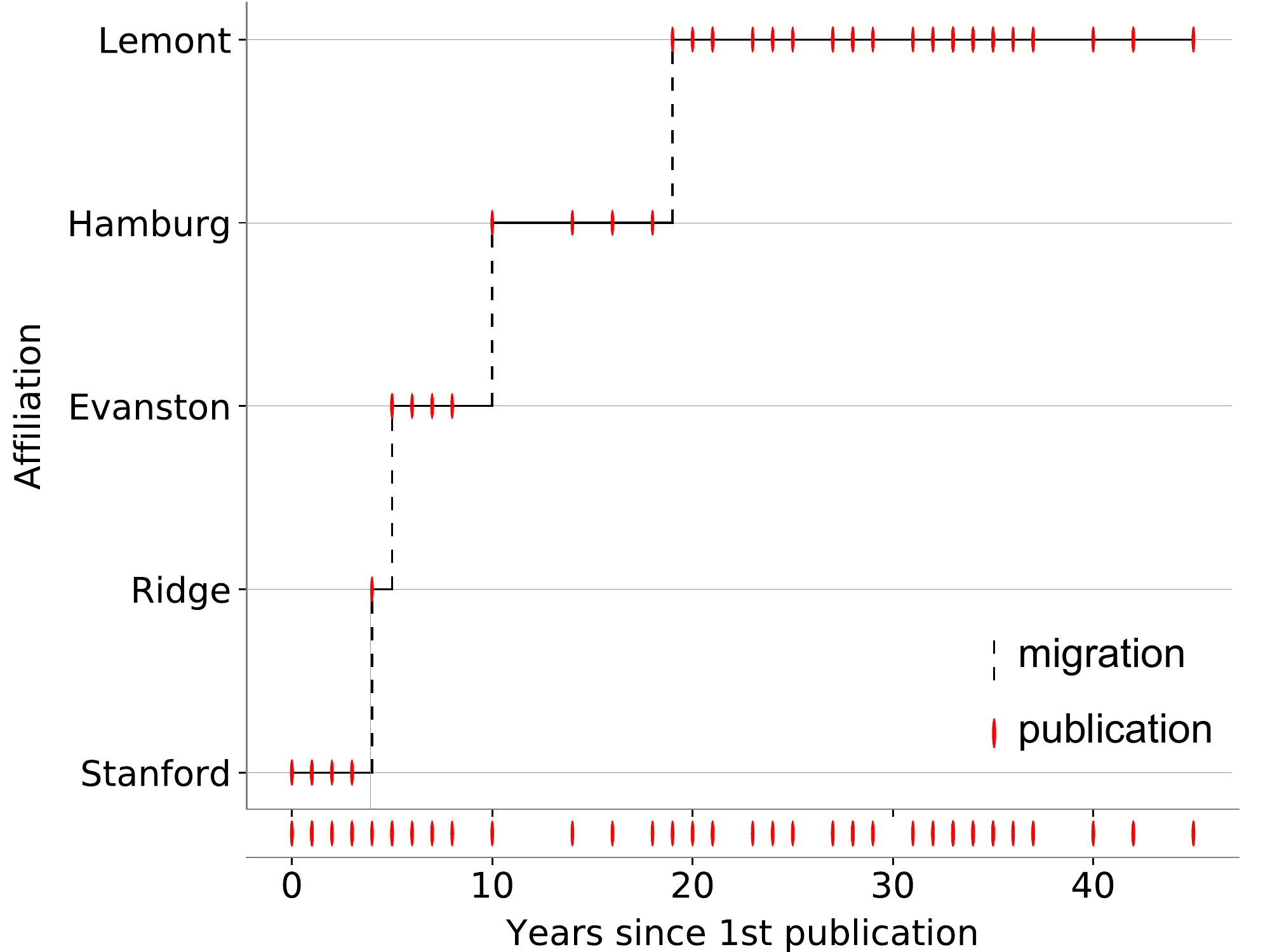}
\caption{The career trajectory of a scientist. Red vertical lines represent publications. Black dashed lines represent moves from an institution to another.}
\label{fig:career_recon}
\end{figure}

Figure \ref{fig:career_recon} shows an illustrative example of a 40-years long career trajectory: the scientist $u$, initially at Stanford University, moves to other four institutions during their career, each migration being detected by the changing of the main affiliation in $u$'s publications.

\subsection{Scientific Profile}
\label{sec:profile}
\newcommand{\vect}[1]{\boldsymbol{#1}}
We define the \emph{scientific profile} of a scientist $u$ during a time window $h$ as the multidimensional feature vector: 

$$\vect{p}^{(h,t)}(u) = [{\underbrace{\textstyle c_1^{(h,t)}, \dots, c_n^{(h,t)}}_{\mathclap{\text{career}}}}, {\underbrace{\textstyle e_1^{(h,t)}, \dots, e_m^{(h,t)}}_{\mathclap{\text{environment}}}}, {\underbrace{\textstyle r_1^{(h,t)}, \dots, r_w^{(h,t)}}_{\mathclap{\text{relationships}}}}],$$ 

\noindent where each element of $\vect{p}^{(h,t)}(u)$ is a feature describing a specific aspect of $u$'s scientific activity during a time window $(t-h ,t)$ of $h$ years ending at time $t$. 

Three macro-aspects compose an individual $u$'s scientific profile: \emph{(i)} their \emph{scientific career}, in terms of $u$'s experience, publications and citations; \emph{(ii)} their \emph{scientific environment}, i.e., the level of production of $u$'s colleagues at $u$'s current institution; and \emph{(iii)} their \emph{scientific relationships}, indicating the working relationships $u$ established with collaborators at external institutions during the $h$ years. Table \ref{tab:features} describes the features composing scientific profile.

\paragraph{Scientific career.} The scientific career 
% indicates how much $u$ has been productive during the $h$ years. 
includes features describing individual characteristics of scientist $u$ in the past $h$ years. 
As proxies of scientific production, we consider the amount of publications the scientist produced and the citations they got during the considered period. Moreover, we estimate $u$'s experience with the number of years since their first publication and define $u$'s scientific mobility using boolean values which represent whether they have or have not changed institutions in the last $h$ years. 

\paragraph{Scientific environment.} A scientist's production shapes, and it is shaped by, the scientific environment where they operate. For this reason we estimate the level of production of $u$'s environment as the number of citations and the number of publications of $u$'s colleagues during the $h$ years. A colleague is a scientist working at the same institution as $u$ at time $t$.
%during the $h$ years. 
Moreover, we consider the differential of citations, i.e., the mean difference between $u$'s citations and their colleagues' citations, as well as  the differential of publications, i.e., the mean difference between $u$'s publications and the publications by their colleagues.

\paragraph{Scientific relationships.} Scientific collaboration is a proven mechanism for promoting excellence in scientific research, as higher collaboration rates are linked to higher citation rates \cite{appelt2015factors, abramo2017relationship,sugimoto2017scientists}.
For this reason we take into account a scientist $u$'s collaborations by estimating the size of $u$'s collaboration circle using three features: the number of institutions $u$ collaborates with during the $h$ years, the total number of distinct collaborators of $u$   
and $u$'s tendency to collaborate with scientist at external institutions (xenophilia), computed as the ratio of external collaborators to the total number of collaborators in the $h$ years.

\begin{table*}[htb]
\centering
\begin{tabular}{|c| c | p{11.5cm} |}
\hline
\textbf{category}&\textbf{variable}& \textbf{description}\\ 
\hline
\multicolumn{1}{|c|}{} & \multicolumn{1}{c|}{publications} & number of papers published in the last $h$ years \\ 
\cline{2-3} 
\multicolumn{1}{|c|}{} & \multicolumn{1}{c|}{citations} & number of citations received in the last $h$ years \\ 
\cline{2-3} 
\multicolumn{1}{|c|}{} & \multicolumn{1}{c|}{experience} & years since the first publication\\ 
\cline{2-3} 
\multicolumn{1}{|c|}{\multirow{-4}{*}{\bf (i) \texttt{career}}} & \multicolumn{1}{c|}{mobility}                                                                & whether she changed institutions in the last $h$ years                                      \\ \hline
\multicolumn{1}{|c|}{}                                & \multicolumn{1}{c|}{\begin{tabular}[c]{@{}c@{}}colleagues'\\ citations\end{tabular}}         & mean number of citations received by colleagues in the last $h$ years                       \\ \cline{2-3} 
\multicolumn{1}{|c|}{}                                & \multicolumn{1}{c|}{\begin{tabular}[c]{@{}c@{}}colleagues'\\ publications\end{tabular}}      & mean number of papers published by colleagues in the last $h$ years                         \\ \cline{2-3} 
\multicolumn{1}{|c|}{}                                & \multicolumn{1}{c|}{\begin{tabular}[c]{@{}c@{}}differential\\ of citations\end{tabular}}     & mean difference between citations and colleagues' citations in the past $h$ years           \\ \cline{2-3} 
\multicolumn{1}{|c|}{\multirow{-8}{*}{\bf (ii) \texttt{environment}}}   & \multicolumn{1}{c|}{\begin{tabular}[c]{@{}c@{}}differential \\ of publications\end{tabular}} & mean difference between publications and the colleagues' publications in the last $h$ years \\ \hline
\multicolumn{1}{|c|}{}                                & \multicolumn{1}{c|}{institutions}                                                            & number of institutions she has collaborated with in the past $h$ years                      \\ \cline{2-3} 
\multicolumn{1}{|c|}{}                                & \multicolumn{1}{c|}{collaborators}                                                           & number of scientists she collaborated with in the last $h$ years                            \\ \cline{2-3} 
\multicolumn{1}{|c|}{\multirow{-4}{*}{\bf (iii) \texttt{relationships}}} & \multicolumn{1}{c|}{xenophilia}                                                              & ratio of external collaborators to total collaborators in the last $h$ years                \\ \hline
\end{tabular}
\caption{The features describing scientific profile with the corresponding description. They are grouped in three macro-categories: (i) career; (ii) environment; and (iii) relationships.}
\label{tab:features}
\end{table*}

\subsection{Problem definition}
\label{sec:problem}
We refer to \emph{next institution prediction} as the task of predicting the institution of a scientist in year $t+1$ given their recent scientific profile up to year $t$. Formally, let $U = \{u_1, \dots, u_z\}$ be a set of $z$ scientists and $I = \{i_1, \dots, i_v\}$ be a set of $v$ institutions. Given a scientist $u \in U$, a history window $h$, a year $t$ and a scientific profile $\vect{p}^{(h,t)}(u)$, the next institution prediction consists of two phases: 
\begin{itemize}
\item {\it Move prediction phase}: predicting whether $u$ will move to a new institution in year $t+1$; 
\item {\it Destination prediction phase}: if $u$ moves, predicting the institution $i \in I$ where $u$ moves to. 
\end{itemize}
We address these phases by the approach detailed in the following section.

\section{Next institution prediction}
\label{sec:methods}

\subsection{Dataset}
Our dataset $D_{\mbox{\tiny APS}}$ consists of all articles published in the American Physical Society (APS) journals from 1950 to 2009. For each article, the date of publication, the author names and the corresponding affiliations are stored. In addition to this, location information (latitude and longitude) is available for every affiliation that appears in $D_{\mbox{\tiny APS}}$. In total, we have over 60,000 scientists, 3,500 institutions and 360,000 articles. 
%\note[Luca]{Do we need here some plots summarizing the structure of the dataset?}

%(actual numbers are 361674, 3732, 61619).

%From the dataset we construct a citation network where a paper $x$ has a link to paper $y$ if $x$ cites $y$.
%: for each paper in the dataset, we have a list of papers cited by the paper, and a list of citations to the paper with the dates at which the citing papers were published.%{\it sorry - I will try to make that sound clearer}
%
\subsection{Computation of career trajectories}
For each scientist $u \in U$ in $D_{\mbox{\tiny APS}}$, we construct a career trajectory $T(u)$ as follows. First, we sort all $u$'s publications by time, from the least recent to the most recent publication, and we build the time-ordered sequence of $u$'s affiliations: 
$A(u) = [(t_1,i_1), \dots, (t_r,i_r)]$, where $r$ is the total number of papers published by $u$, $t_j$ is the year of publication of $u$'s $j$-th paper and $i_j$ is $u$'s affiliation on paper $j$ ($1 \leq j \leq r$). 
Note that a scientist may specify more than one affiliation in a publication. We disambiguate these cases using the first affiliation reported by the scientist, as suggested in the literature~\cite{deville2014career}. 
We then initialize $T(u) = \langle (t_1, i_1) \rangle$ and add tuple $(t_j, i_j)$, with $j = 2, \dots, r$, to $T(u)$ if $i_j \neq i_{j - 1}$, i.e., if the institution associated with publication $j$ is different from the institution associated with the previous publication $j-1$. Algorithm \ref{alg:career_traj} shows the pseudo-code of the algorithm to build $u$'s career trajectory from their list of publications.

\begin{algorithm}
 \small
\DontPrintSemicolon
\SetKwInOut{Input}{input}
\SetKwInOut{Output}{output}
    \Input{$L_u=\{a_1, \dots, a_r\}$, list of $u$'s publications\\
    }
    \Output{$T(u)$, career trajectory of $u$}
\setcounter{AlgoLine}{0}    
    \SetKwFunction{sortList}{sortList}
    \SetKwFunction{extractMainInstitution}{extractMainInstitution}
    \SetKwFunction{append}{append}
    \DontPrintSemicolon
    \normalsize
    \nl $L_u^* = \extractMainInstitution(L_u)$ \tcp{extract tha main institution from every paper}
    \nl $l_0 = L_u^*[0]$\;
    \nl $t, i = l_0.t, l_0.i$\;
    \nl $T(u).\append( (t, i) )$\;
    \nl \ForAll{$j \in \{1, \dots, r\}$}{
    \nl $l_j, l_{j - 1} = L_u^*[j], L_u^*[j - 1]$\;
    \nl $t, i = l_j.t, l_j.i$\;
    \If{$i \neq l_{j-1}.i$}
    {
    \nl $T(u).\append((t,i))$\;
    }
     }
    \KwRet $T(u)$
    \small
\caption{Algorithm to build a scientist's career trajectory from their list of publications.}
\label{alg:career_traj}
\end{algorithm}

\subsection{Computation of scientific profile}
Given time $t$ and history window $h$, from $D_{\mbox{\tiny APS}}$ we compute the career, environment and relationships features for the scientist's activity in the $h$ preceding years in the following way.
The number of publications by a scientist $u$ is given by the total number of papers in $D_{\mbox{\tiny APS}}$ for which $u$ is an author and the publication date falls within the period $(t-h, t)$. We compute the number of citations of $u$ as the sum of citations to all papers in $D_{\mbox{\tiny APS}}$ for which $u$ is an author and for which the citing paper is published in the period $(t-h, t)$. The experience of $u$ is computed as the difference $t -t_1$, where $t_1$ is the time of $u$'s first publication in $D_{\mbox{\tiny APS}}$. Finally, the mobility of $u$ relates to how recently they have changed institution: if $u$ has moved within the period $(t-h, t)$, the mobility feature has value $1$, if they have not moved the feature has value $0$, if at time $t$ they are at their first institution (i.e., the only institution they have been affiliated with so far in the dataset) they are assigned a value of $-1$.

To compute the environmental features we define the \emph{colleagues} (or peers) of $u$ as all the scientists who publish a paper during the period $(t-h, t)$ that are affiliated with $u$'s institution at time $t$. For each colleague we compute their publications and citations as described above. The colleagues' citations and publications features are then computed as the mean number of citations and the mean number of publications for all colleagues, respectively. The differential of citations is determined by taking the difference between the scientist's citations and each peer's citations individually, and then taking the mean of these differences. The differential of citations is determined in an identical way: the mean difference between the number of publications made by the scientist and each peer.

To compute the relationships features we define a \emph{collaborator} as a scientist who is a co-author of $u$ for at least one paper published in the period ($t-h$,$t$). The collaborators feature is then the number of distinct collaborators in the list of co-authors and the institutions feature is the number of distinct affiliations. We compute xenophilia as the ratio between the number of collaborators at external institutions divided by the total number of collaborators. 

%For our second model, where you go if you go, we include additional features which compare a scientist to the scientists at possible destinations. To do this, we define the new peers: all scientists whom publish a paper during the period ($t-h$,$t$) that affiliates them to the destination institution. For these new peers we compute the publications, citations, differential of publications and differential of citations as described above.

\subsection{Phase \#1: Move Prediction}
The move prediction phase aims at predicting if a scientist will migrate the next year given their recent scientific profile. Given the set of scientists $U$, for each $u \in U$ we compute the feature vector $\vect{p}^{(h,t)}(u)$ that describes her scientific profile between years $t-h$ and $t$. Since the features in $\vect{p}^{(h,t)}(u)$ have different ranges of values and distributions, we standardize them by computing their quantiles with respect to each feature of the other scientists in $U$. By doing this, each element in $\vect{p}^{(h,t)}(u)$ lies between $0$ and $1$ and represents how the features of $u$ compare to all other scientists $u \in U$.

We then assign $u$ to a label $m(u) \in \{0, 1\}$, where 1 indicates that $u$ will migrate the next year, and 0 that $u$ will not migrate. From the scientists' feature vectors we construct a dataset of examples $\vect{p}^{(h,t)} = \{\vect{p}^{(h,t)}(u) | u \in U\}$ each associated with the corresponding label in $m = \{m(u) | u \in U\}$. 
For examples where $m(u)=1$, we use the move events of all $60,000$ scientists: all values of $t$ for which the affiliation of $u$ in year $t+1$ is different from the affiliation of $u$ in year $t$.
%), defining $t$ as the year before $u$ moved. 
%
For examples where $m(u)=0$, for each $u \in U$ we generate a random value $t$ between $u$'s first and last publications in $D_{\mbox{\tiny APS}}$ until $t+1$ is not a year in which $u$ moved. We repeat this process $3$ times to ensure that our classes, $m(u) =0$ and $m(u)=1$ are approximately balanced. 
In total, our dataset contains $290,000$ examples: $140,000$ examples for which $m(u) =1$ and $150,000$ examples for which $m(u) =0$.

To investigate to what extent we can predict a future migration given a scientist's history, for each value $h = 1, \dots, 10$ we train two predictors $L_h$ and $T_h$ on the dataset $\vect{p}^{(h,t)}$ and the associated labels $m$, where $L_h$ is a logit and $T_h$ is a decision tree.\footnote{We use the Python package \texttt{scikit-learn} \cite{sklearn}.} We evaluate the predictors with 10-fold cross-validation and investigate the goodness of predictions -- in terms of accuracy, recall, precision, F1-score and AUC -- varying $h = 1, \dots, 10$. Table \ref{tab:m1_results} shows the goodness of prediction of the best tree and the best logit. 
%\todo[Charlotte]{Insert here the new results of the models varying $h=1, \dots, 10$.}
%We find that $h_{best} = 10$ for both $T_h^k$ and $L_h^k$, while $k_{best} = 3$ for $T_h^k$ and $k_{best} = 9$ for $L_h^k$.

\begin{table}[htb]\centering
\def\arraystretch{1.5}
\begin{tabular}{|c | r || r |r |r |r |r |}
\hline
 \bf model & $h_{\mbox{\footnotesize best}}$ & \bf ACC & \bf recall & \bf prec & \bf F1 & \bf AUC\\
\hline
 tree & 10 & \bf 0.84 & \bf 0.81 & 0.85 & \bf 0.83 & \bf 0.84\\
 \hline
 logit & 5 & 0.82 & 0.72 & \bf 0.87 & 0.79 & 0.81 \\
 \hline
 \hline
\em dummy & - & 0.50 & 0.47 & 0.47 & 0.47 & 0.50 \\
 \hline
 \end{tabular}
 \caption{Predictive performance of the best tree ($T_{10}$) and the best logit ($L_5$), compared with a baseline classifier (dummy). The models are evaluated with a 10-fold cross-validation, and the goodness of predictions are measures in terms of accuracy (ACC), recall, precision (prec), F1-score (F1) and Aurea Under the Curve (AUC).}
 \label{tab:m1_results}
 \end{table}

\begin{figure}[htb]\centering
\includegraphics[scale=0.4]{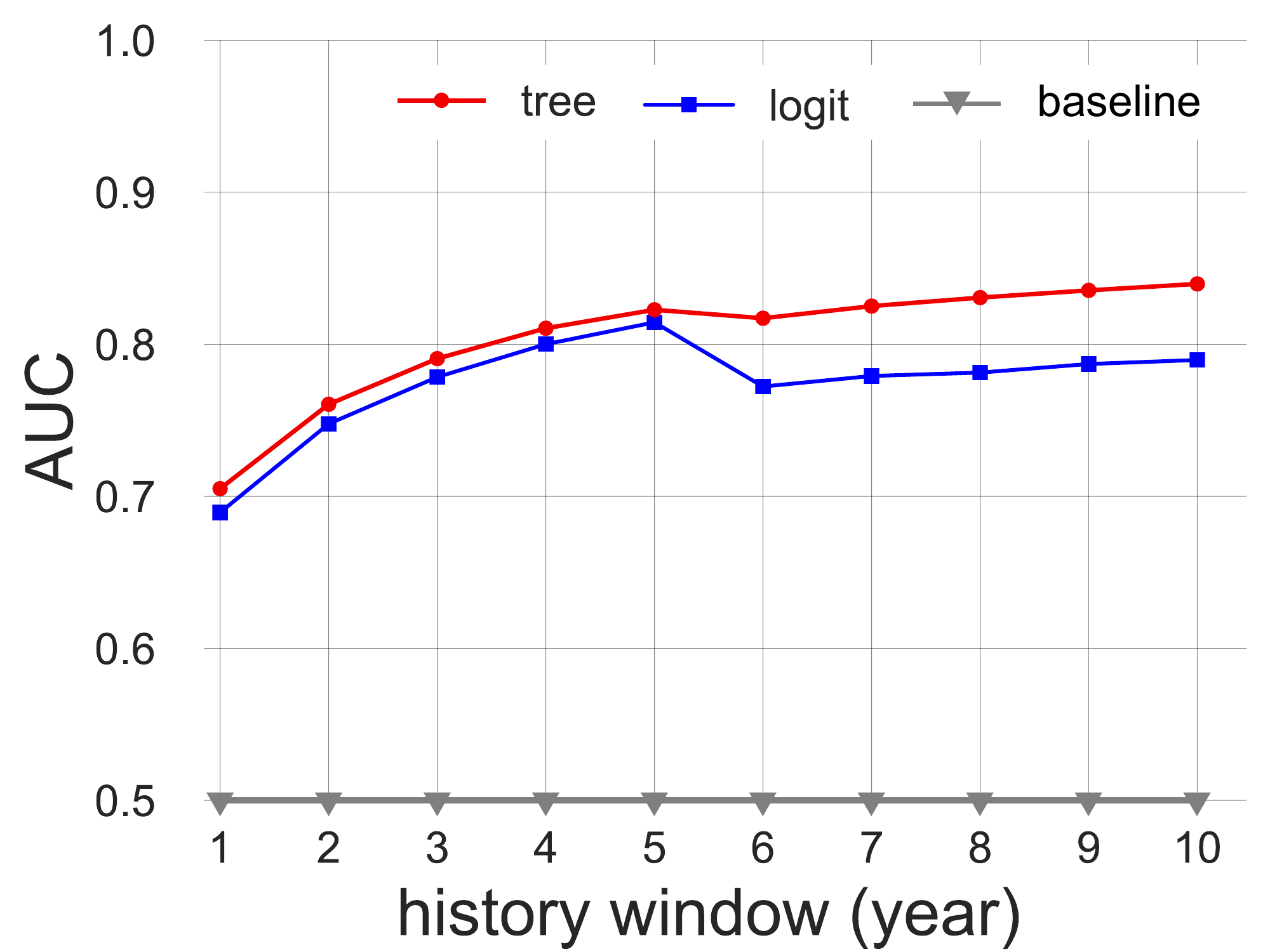}
\caption{AUC score of predictors $T_h$ and $L_h$, for $h = 1, \dots, 10$. The predictions improve as more years are considered up to $h=5$. For $h > 5$, the predictive performance stabilizes. The baseline classifier has $AUC = 0.5$.}
\label{fig:auc}
\end{figure}

We find that $h_{\mbox{\footnotesize best}}=10$ for $T_h$ and $h_{\mbox{\footnotesize best}}=5$ for $L_h$. Both predictors perform better than a baseline classifier which generates predictions based upon the class distribution of the training set. 
Figure \ref{fig:auc} shows how the AUC score changes with the size of the history window for $T_h$ and $L_h$, $h=1,\dots,10$. We observe that: (i) the tree performs slightly better than the logit; (ii) both curves stabilise at $h=5$ suggesting that a history window $h>5$ adds no additional predictive power to either model; (iii) both curves decrease slightly between $h=5$ and $h=6$. The decrease may be explained by a change in feature values due to, for example, the average length of a PhD or tenure track position being approximately 5 years. In other words, the last five years of a scientist provide with sufficient information to predict, with reasonable accuracy, whether they will move or not.

%for $h=1,\dots,10$. This indicates that the AUC of both models stabilises at $h=5$ suggesting that a history window $h>5$ adds no additional predictive power to either model. The drop in $AUC$ at $h=6$ may be explained by the 

\begin{figure}[htb]\centering
\includegraphics[scale=0.325]{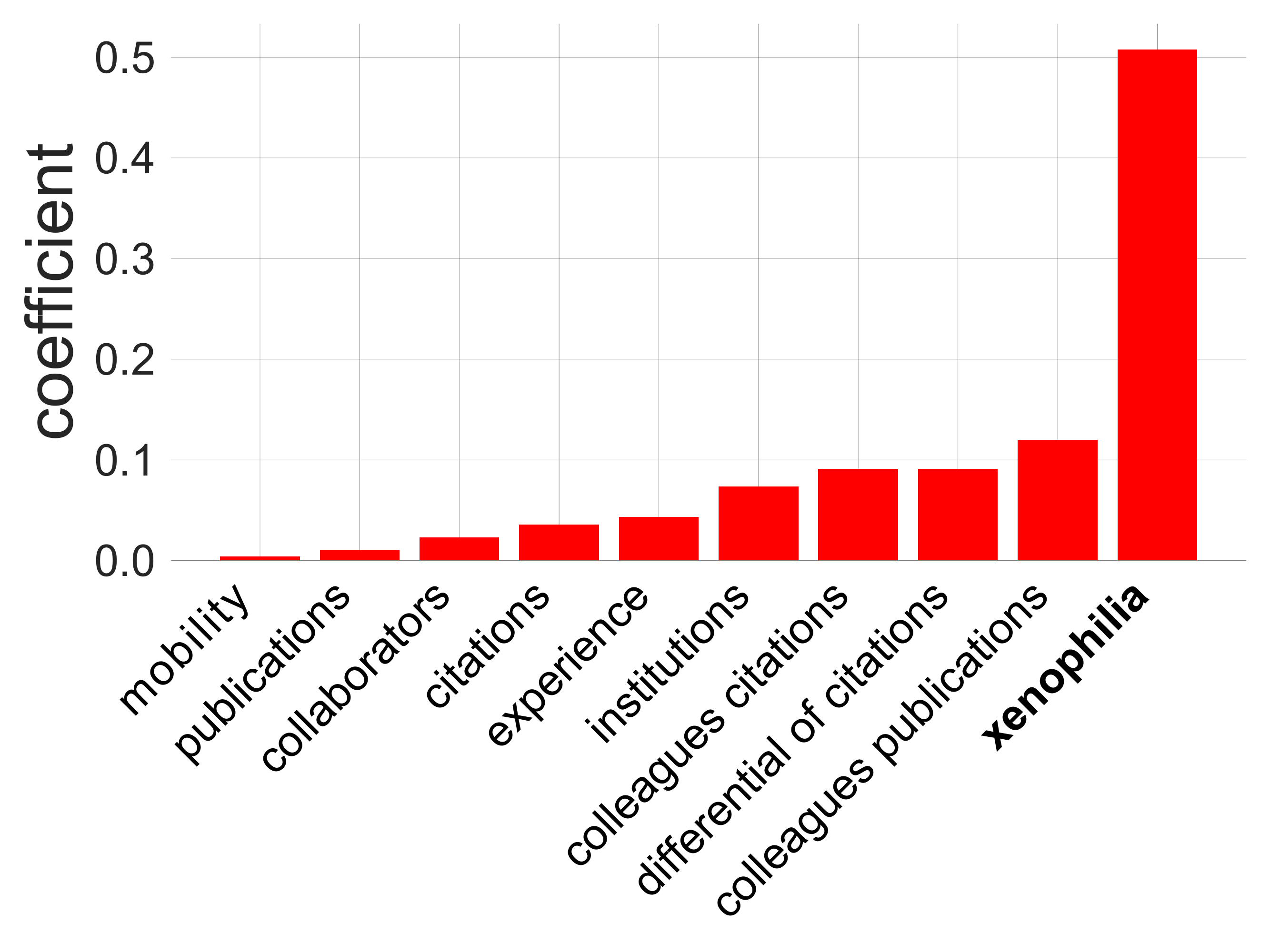}
\caption{
%{\color{red}[Use the same feature names of Table 1]} 
Coefficients resulting from model $T_{10}$. The ratio of external collaborators (xenophilia) is the strongest predictor of future migration.}
\label{fig:weights_M1_h5_k1}
\end{figure}

\begin{figure}[htb]\centering
\includegraphics[scale=0.4]{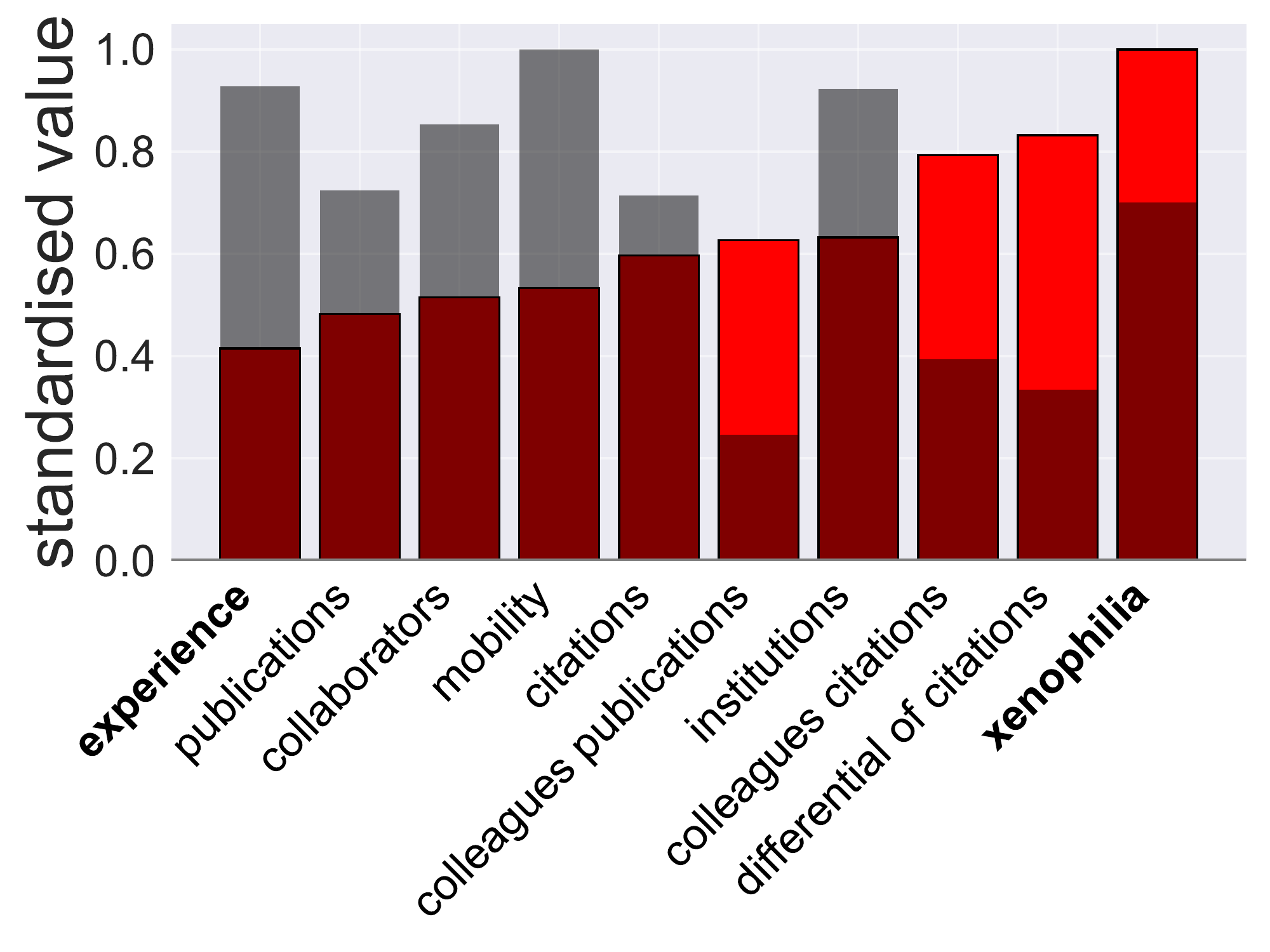}
\caption{Feature values for a scientist predicted to move by model $T_{10}$ (red) and a scientist predicted to stay (grey). 
%{\color{red}[Use the same feature names of Table 1]} 
}
\label{fig:compare_m1}
\end{figure}

Figure \ref{fig:weights_M1_h5_k1} shows the feature importances resulting from $T_{10}$, indicating that xenophilia (i.e., the ratio of external collaborators to total collaborators) is the strongest predictor of the probability to migrate. This means that scientists with a strong propensity to collaborate with other scientists in external institutions are more likely to migrate. 
Figure \ref{fig:compare_m1} shows the standardised features of a scientist correctly predicted to move (red bars) and a scientist correctly predicted to stay (grey bars) using model $T_{10}$. We observe that the scientist who moves has high levels of xenophilia compared to the scientist who does not move. In contrast, the scientist who does not move scores highly against other scientists for features, such as mobility and collaborators, that are of little importance according to Figure \ref{fig:weights_M1_h5_k1}.

%We also consider a different definition of future window, where we try to predict whether a scientist will migrate at least ones within the next $k$ years. Figure \ref{fig:heatmap_auc}c-d shows how AUC changes with the history window $h$ and the future window $k* = \in \{1, 2, 5, 10\}$. We find that there is a slightly increasing trend with $h$ and that the higher the future window $k*$ the higher the AUC tends to be in the decision tree. 

\subsection{Phase \#2: Destination prediction}

The destination prediction phase aims at predicting the institution of destination of a scientist that will move. 
The task of estimating the probability that a scientist will relocate to a given institution can be interpreted as a classification problem with as many classes as institutions. 
The state-of-the-art model to estimate mobility and migration flows is the gravity model \cite{RefWorks:134,jung2008,Pappalardo2016934}, which is a multinomial logistic regression based on distance and size (population) of locations. 
Specifically, the features of a traditional singly-constrained gravity model are the size of the potential destination, estimated here as the logarithm of the number of scientists affiliated to the destination institution, and the logarithm of the distance between the scientist's current institution and the potential destination.
We compare the performance of the traditional gravity model with a \emph{social-gravity model}, an extended model that includes additional indicators of quality and social interactions with the potential destination. 
The new social and quality features are: \emph{(i)} the fraction of collaborators at the potential destination; \emph{(ii)} the average number of papers published by the scientists at the potential destination; \emph{(iii)} the average number of citations received by the scientists at the potential destination. We compute these features using $h=5$. 

We train the model maximising its likelihood using stochastic gradient descent over the train set $U_{\mbox{\footnotesize train}} \subset U$, which contains 200 scientists selected at random from $U$. 
In order to reduce the computational cost of the optimisation, which can be quite high when the total number of destination locations is over $10^3$, we consider an approximation of the likelihood computed considering a subset of 100 potential destinations, as proposed in~\cite{li2017nemo}. The subset of potential destinations is extracted for each move of each scientist in $U_{\mbox{\footnotesize train}}$. This subset of potential destinations always includes the true destination, while the other locations are randomly selected with a probability that increases linearly with their sizes and slowly decreases with the distances from the origin location. This ensures that the most relevant potential destinations, i.e., the larger and closer to the origin, are included in the likelihood's estimate. Numerical tests show that the optimal size of the subset of potential destinations is 100 locations, i.e., considering more than 100 potential destinations does not significantly improve the model's performance. 

\begin{figure}[htb]\centering
\includegraphics[scale=0.4]{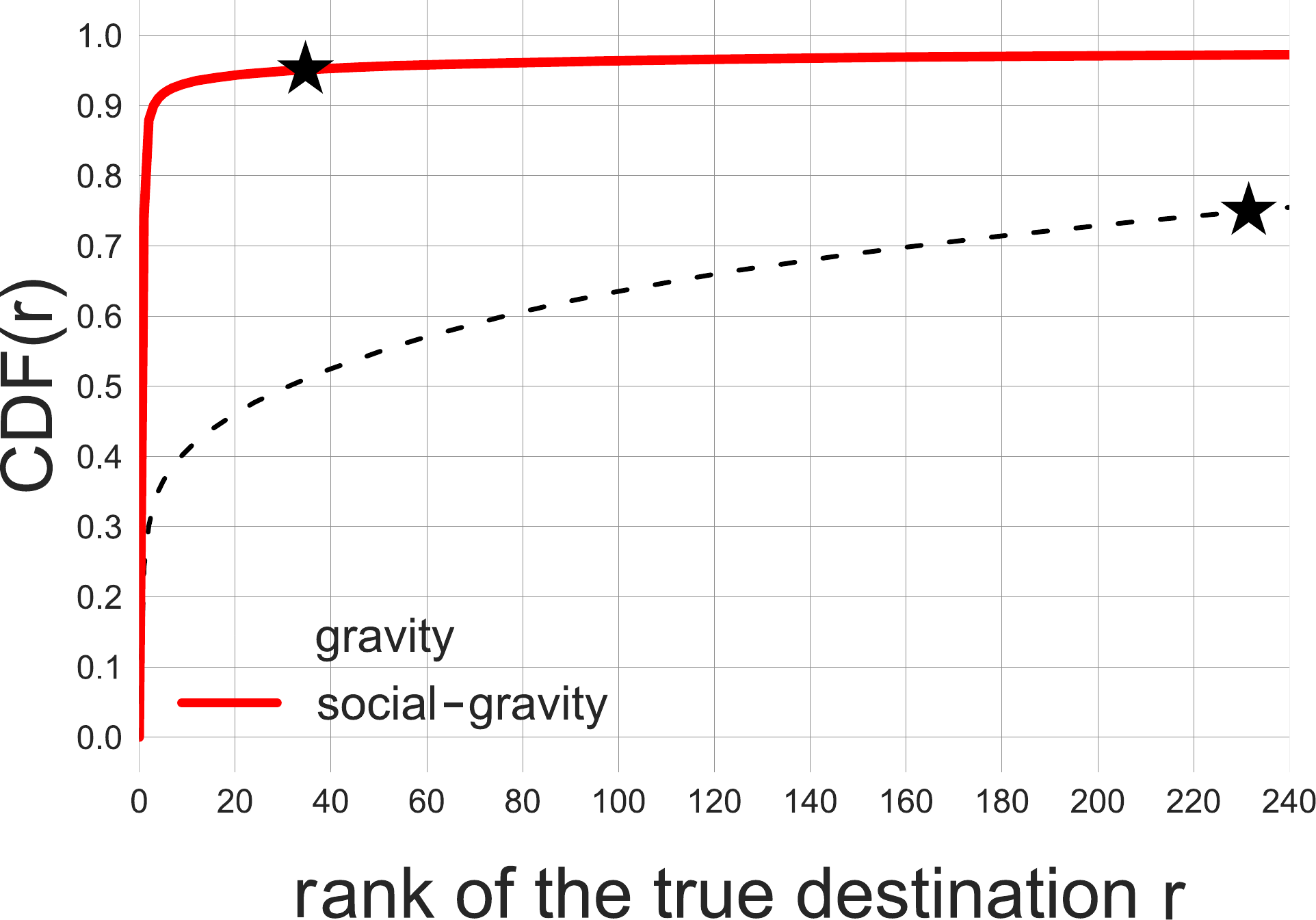}
\caption{{CDF of the ranks of scientists' true destinations according to the original gravity model (dashed black) and the social-gravity model (red). Black stars represent the mean rank: we find that $\bar{r} = 231$ for the original gravity model and $\bar{r} = 34$ for the social-gravity model.}}
\label{fig:ranks}
\end{figure}

The model's performance is evaluated considering all the remaining scientists in $U$, i.e., $U_{\mbox{\footnotesize test}} = U \setminus U_{\mbox{\footnotesize train}}$. For each scientist $u \in U_{\mbox{\footnotesize test}}$ we compute the probabilities, $P(u,i)$, to relocate to any institution $i \in I$ according to the model. We then sort all institutions in decreasing order of $P(u,i)$ and define the rank of each institution so that the model's top prediction has the largest $P(u,i)$ and rank 1. We then consider the rank of the scientist's actual destination, $r_u$, and we use it to compute two statistics: (1) the cumulative distribution function (CDF) of the ranks $r$ of all the moves of all scientists in $U_{\mbox{\footnotesize test}}$ (Figure~\ref{fig:ranks}); and (2) the Mean Percentile Ranking (MPR)~\cite{li2017nemo} defined as: 

$$MPR = \frac{1}{|U_{\mbox{\footnotesize test}}|} \sum_{u \in U_{\mbox{\footnotesize test}}} \frac{r_u}{|I|},$$ 

\noindent where $r_u$ is the rank of scientist $u$'s actual destination. 

Results show that including information on the collaboration network significantly improves the model's performance. 
In particular, the social-gravity model that includes quality and social information has CDF$(r{=}10)=0.93$, i.e., for 93\% of the scientists in $U_{\mbox{\footnotesize test}}$ the real destination is among the top 10 model's predictions. 
This is considerably better than the original gravity model without quality and social information, which has CDF$(r{=}10)=0.41$, i.e., only for 41\% of the scientists the real destination is among the top 10 model's predictions (see Figure \ref{fig:ranks}). This result, that a social-gravity model which incorporates social information is superior, is further supported by the MPR: $MPR_{\mbox{\footnotesize social}} = 0.009$ while $MPR_{\mbox{\footnotesize gravity}} = 0.062$, corresponding to an error reduction of 85\%. These results are summarised in Table \ref{tab:m2_results}.

\begin{table}[htb]\centering
\def\arraystretch{1.5}
\begin{tabular}{|c || r | r |r |}
\hline
 \bf model & \bf MPR & \bf CDF (r=10)& \bf $\bar{r}$\\
\hline
\hline
 gravity & 0.062 &  41\% & 231 \\
 \hline
\bf \em social-gravity & \bf 0.009 & \bf 93\% & \bf 34  \\
 \hline

 \end{tabular}
 \caption{Performance comparison between the original gravity model and the social-gravity model}
 \label{tab:m2_results}
 \end{table}
 
\begin{figure}
\centering     
\includegraphics[scale=0.52]{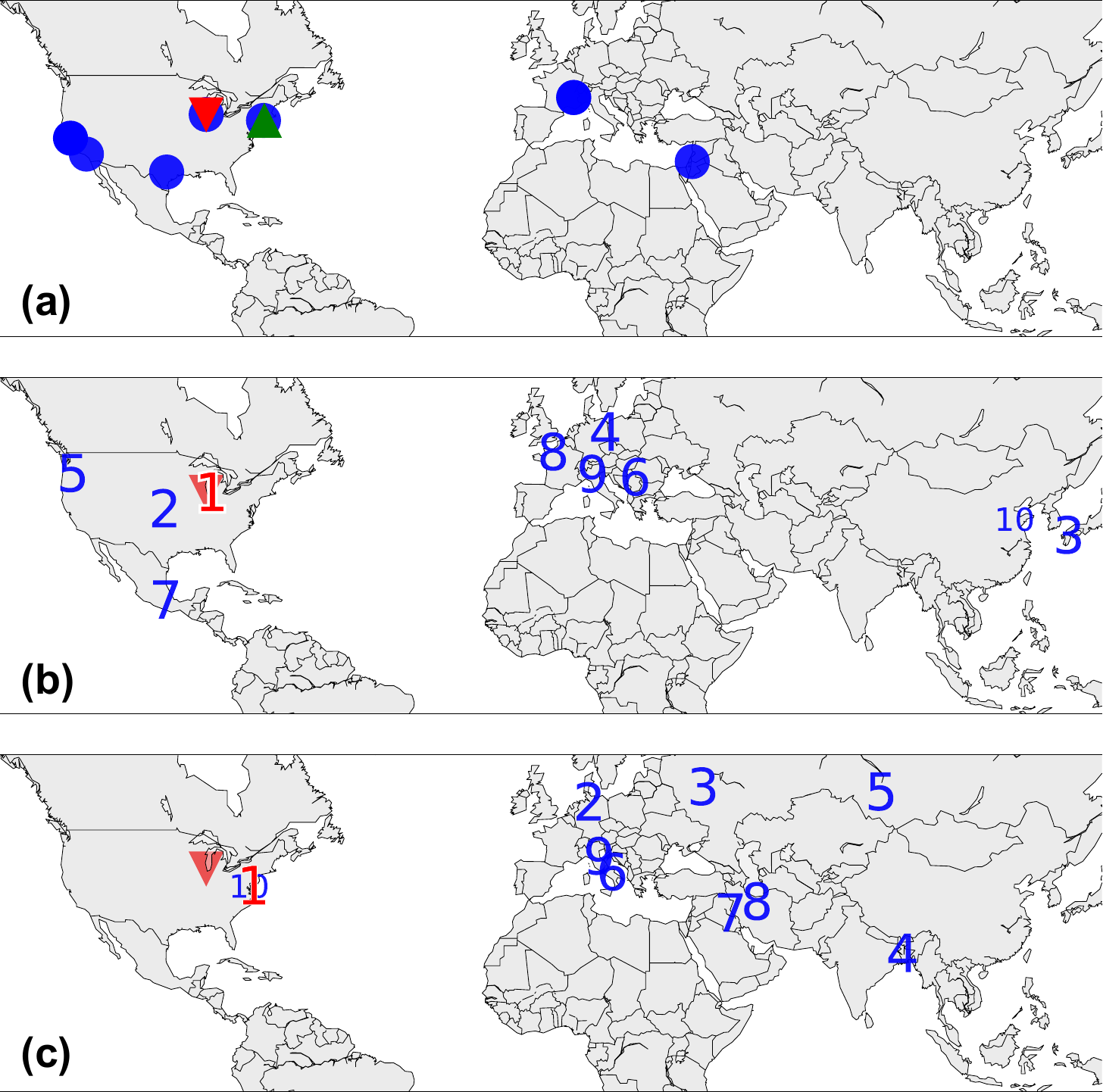}
\caption{Results from the gravity-based model for the scientist in Figure \ref{fig:career_recon}'s move between Ridge and Evanston. (a) Green triangle represents origin, red triangle represents destination. Blue circles represent the locations of collaborators. (b) Ranks and locations of the 10 most probable destinations according to the social-gravity model. The true destination has a rank of 1: the model correctly predicted the scientist's next career move. (c) Ranks and locations of the 10 most probable destinations according to the original gravity model. Here the destination with rank 1 is not the actual destination.}
\label{fig:maps}
\end{figure}

Figure \ref{fig:maps} shows the international collaborations and the predicted destinations for scientist $u \in U$ whose relocations are shown in Figure \ref{fig:career_recon}. In Figure \ref{fig:maps}a we see that the institutions of $u$'s collaborators, plotted as blue circles, are spread around the world, reflecting $u$'s high xenophilia. In Figure \ref{fig:maps}b, the institutions are ranked according to the predictions of the social-gravity model. We observe that $u$'s true destination (the red triangle) has rank 1, indicating that the social-gravity model correctly predicts the scientist's next career move. This is in contrast to Figure \ref{fig:maps}c, where institutions are ranked according to the original gravity model and we observe that the highest ranked institution is not the true destination, rather it is an institution in close proximity to the origin (the green triangle).

%\subsection{Phase \#3: Nationality Analysis}

Are the migration decisions of scientists influenced by the place where they reside? We investigate this aspect by selecting scientists residing in Europe, $u \in E$, where $E \subset U$, and the United States of America, $u \in A$, $A \subset U$. 
We find that, if $U_{\mbox{\footnotesize train}} = E$ and $U_{\mbox{\footnotesize test}} = A$, for both prediction phases there is no significant change in the performance of the models. This is also confirmed for the case $U_{\mbox{\footnotesize train}} = A$, $U_{\mbox{\footnotesize test}} = E$. This means that there are no significant differences behind the decision making of scientists originating in Europe and the United States. In other words, our two-phase approach is applicable on both global scientific migration and scientific migration from a single continent.

\section{Conclusions and future works}
\label{sec:conclusions}

In this paper, we proposed an efficient solution to the problem of predicting the future institution of a scientist. In the first \emph{move prediction phase}, we used data mining to predict whether or not a scientist will migrate on the basis of the quality of their career, environment and collaborations. Given the good accuracy achieved in solving the first phase, we moved to the second, and more challenging, \emph{destination prediction phase}. Experiments showed that our \emph{social-gravity model}, obtained by injecting information about a scientist's collaboration network into the traditional gravity model, achieves a prediction error which is up to 85\% lower than state-of-the-art approaches.

Our results provide us with a deeper understanding of the factors influencing a scientist's decision to relocate. We discovered that features associated with scientific collaboration are a key factor in the decision to move, highlighting the importance of social interactions in human activities. Previous studies in human mobility have demonstrated that information on individual human mobility can be used to improve link prediction in social networks \cite{cho2011friendship,wang2011human}. Our results suggests that the opposite is also true: information on individual social connections can be used to significantly increase the predictive power of migration models. 

Our work paves the road to many research lines. For example it would be interesting to generalize our approach to other classes of high-skilled individuals, such as senior managers \cite{iredale2016high} or soccer players \cite{liu2016anatomy}. As the relocation of those individuals strongly affects the success of both the origin and destination companies, predicting their relocation decisions can have wide economic consequences in the companies' revenues and future profits. We plan to exploit the proposed framework for the creation of a human migration model for the general population. In this context, we can use mobile phone data and social media data to describe individual relocations and social relationships, respectively, and construct a social-gravity model for human migration.

\section*{Acknowledgments}
Luca Pappalardo and Alina S\^irbu were supported by the  European Commission through the Horizon2020 European project "SoBigData Research Infrastructure --- Big Data and Social Mining Ecosystem" (grant agreement 654024). Filippo Simini is supported by EPSRC First Grant EP/P012906/1. We thank Daniele Fadda for his support on data visualisation.

\bibliographystyle{abbrv}
\bibliography{biblio.bib} 

\end{document}